\documentclass[sigconf, authorversion]{acmart}

\usepackage{subcaption}

\copyrightyear{2026}
\acmYear{2026}
\acmConference[Preprint]{}{March 2025}{}
\acmBooktitle{Preprint} 
\acmDOI{10.1145/3772363.3798846}
\acmISBN{979-8-4007-2281-3/2026/04}

\begin{document}

\title{Designing for Adolescent Voice in Health Decisions: Embodied Conversational Agents for HPV Vaccination}

\author{Ian Steenstra}
\affiliation{%
    \institution{Northeastern University}
    \city{Boston, MA}
    \country{USA}
}
\email{steenstra.i@northeastern.edu}

\author{Neha Patkar}
\affiliation{%
    \institution{Tufts Medical Center}
    \city{Boston, MA}
    \country{USA}
}
\email{neha.patkar@tuftsmedicine.org}

\author{Rebecca B. Perkins}
\affiliation{%
    \institution{Tufts Medical Center}
    \city{Boston, MA}
    \country{USA}
}
\email{rebecca.perkins1@tuftsmedicine.org}

\author{Michael K. Paasche-Orlow}
\affiliation{%
    \institution{Tufts Medical Center}
    \city{Boston, MA}
    \country{USA}
}
\email{michael.paasche-orlow@tuftsmedicine.org}

\author{Timothy Bickmore}
\affiliation{%
    \institution{Northeastern University}
    \city{Boston, MA}
    \country{USA}
}
\email{t.bickmore@northeastern.edu}

\renewcommand{\shortauthors}{Steenstra et al.}

\begin{abstract}
Adolescents are directly affected by preventive health decisions such as vaccination, yet their perspectives are rarely solicited or supported. Most digital interventions for Human Papillomavirus (HPV) vaccination are designed exclusively for parents, implicitly treating adolescents as passive recipients rather than stakeholders with agency.  We present the design and evaluation of a mobile intervention that gives adolescents a voice in HPV vaccination decisions alongside their parents. The system uses embodied conversational agents tailored to each audience: parents interact with an animated physician using education and motivational interviewing techniques, while adolescents can choose between an age-appropriate doctor or a narrative fantasy game that conveys HPV facts through play. We report findings from a clinic-based pilot study with 21 parent–adolescent dyads. Results indicate high satisfaction across both audiences, improved HPV knowledge, and increased intent to vaccinate.  We discuss design implications for supporting adolescent participation, choice, and agency in decisions about their health.
\end{abstract}



\begin{teaserfigure}
  \centering
  \begin{subfigure}[b]{0.18\textwidth}
    \includegraphics[width=\textwidth]{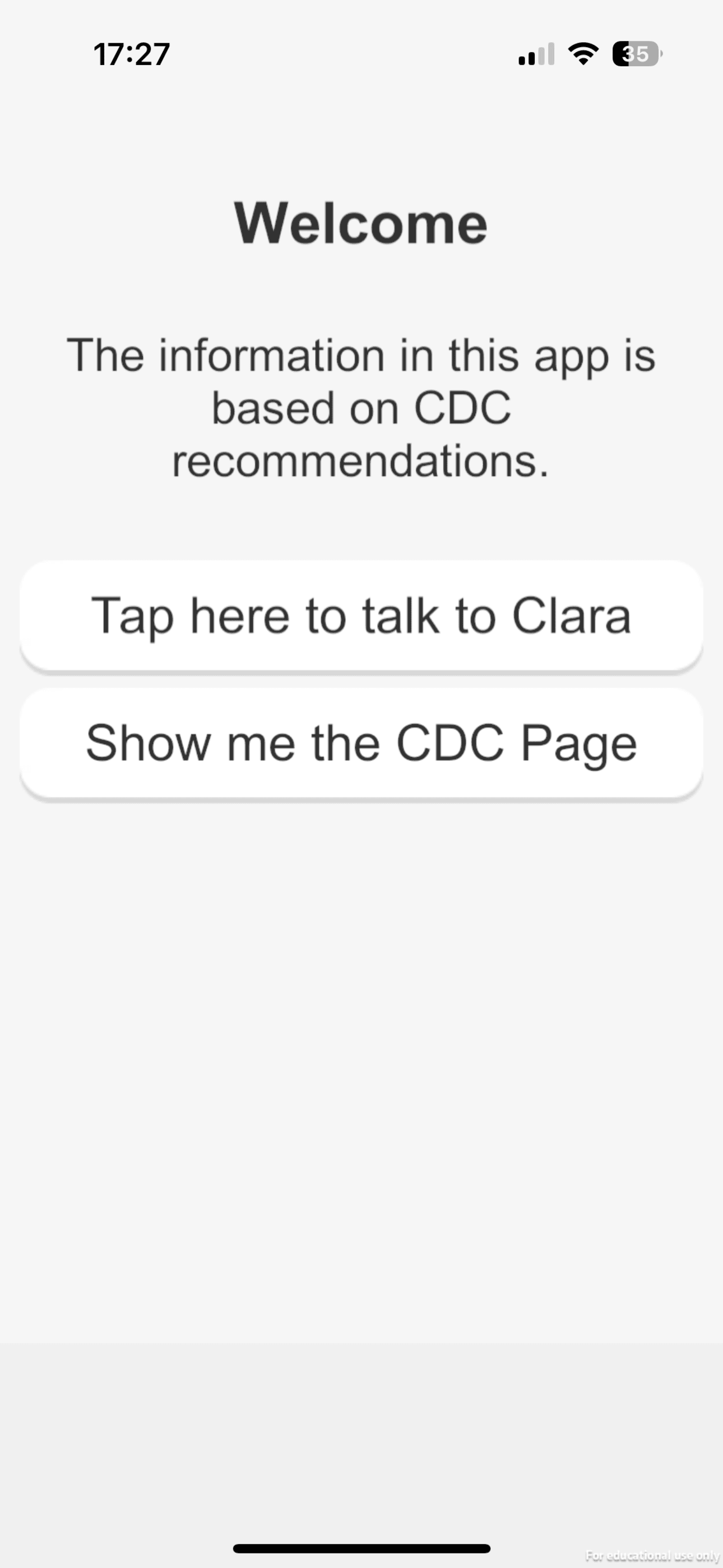} 
    \caption{Welcome Screen}
    \label{fig:welcome}
  \end{subfigure}
  \hfill
  \begin{subfigure}[b]{0.18\textwidth}
    \includegraphics[width=\textwidth]{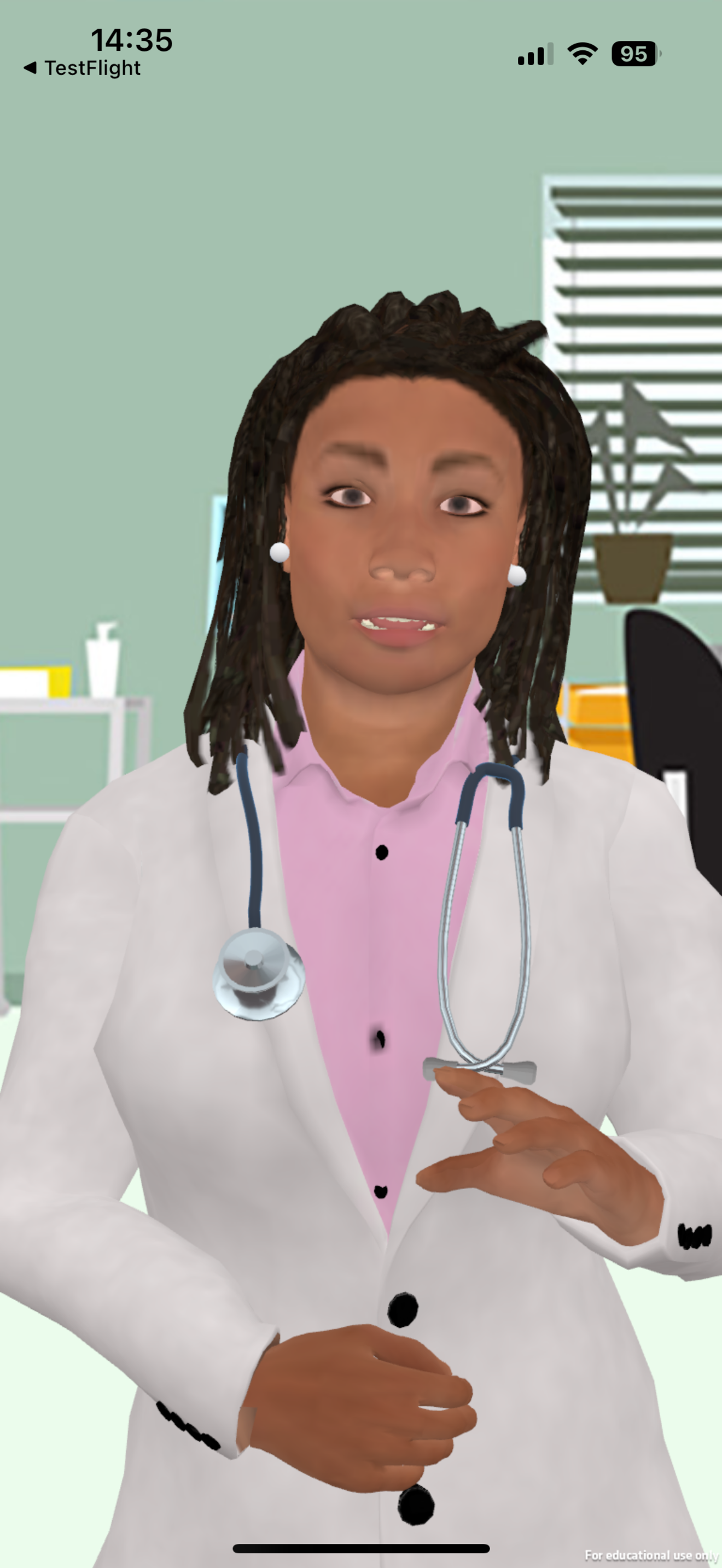}
    \caption{Dr. Clara}
    \label{fig:dr-clara}
  \end{subfigure}
  \hfill
  \begin{subfigure}[b]{0.18\textwidth}
    \includegraphics[width=\textwidth]{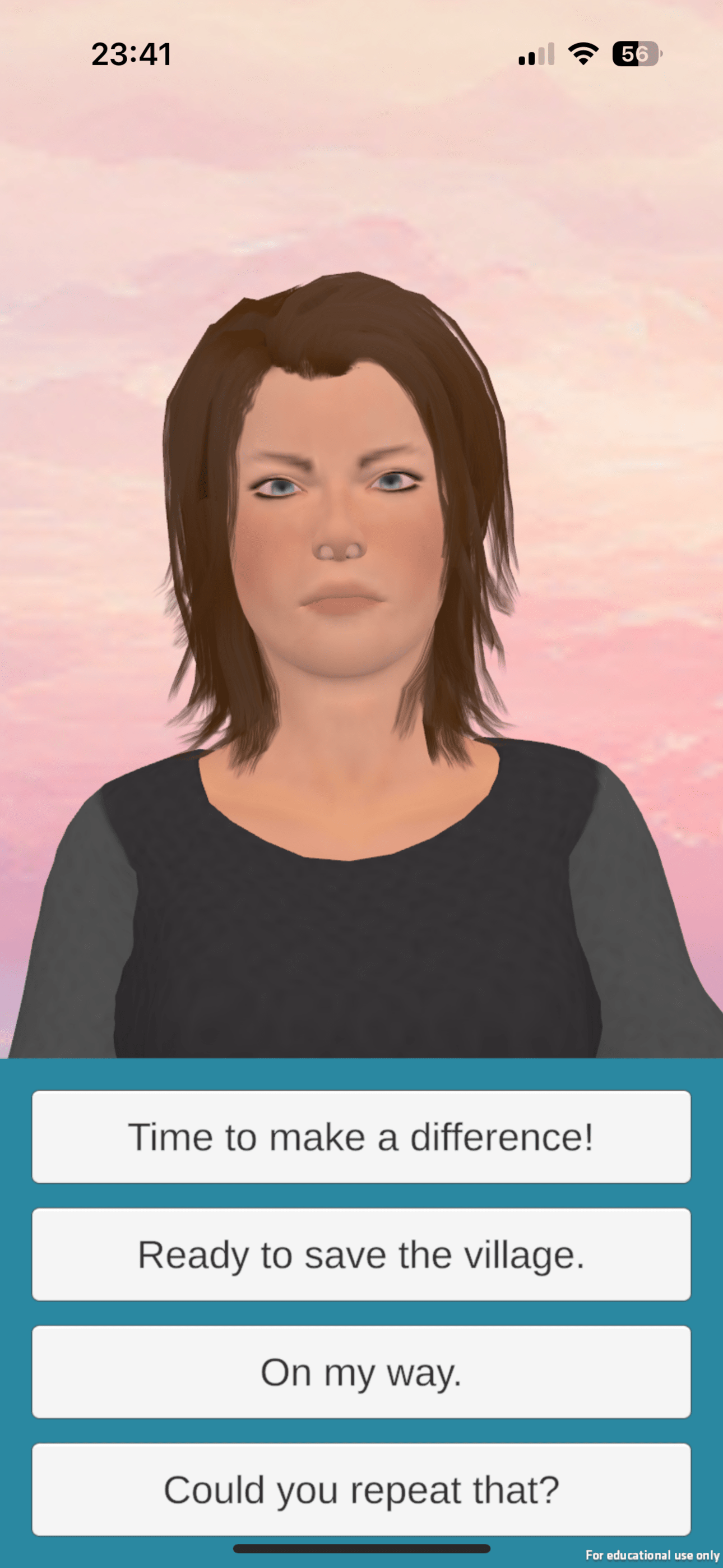} 
    \caption{Game Narrator}
    \label{fig:narrator}
  \end{subfigure}
  \hfill
  \begin{subfigure}[b]{0.18\textwidth}
    \includegraphics[width=\textwidth]{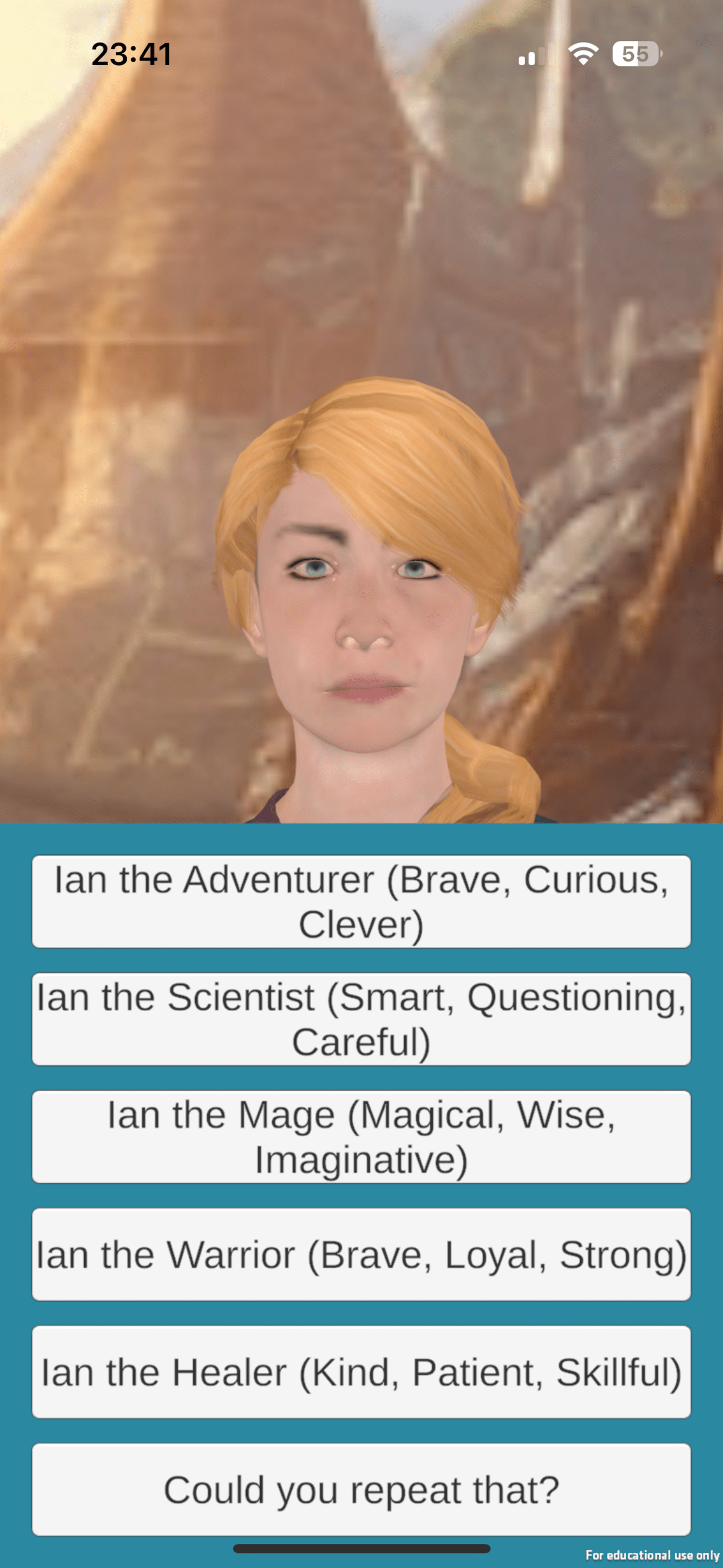} 
    \caption{Character Selection}
    \label{fig:role-select}
  \end{subfigure}
  \hfill
  \begin{subfigure}[b]{0.18\textwidth}
    \includegraphics[width=\textwidth]{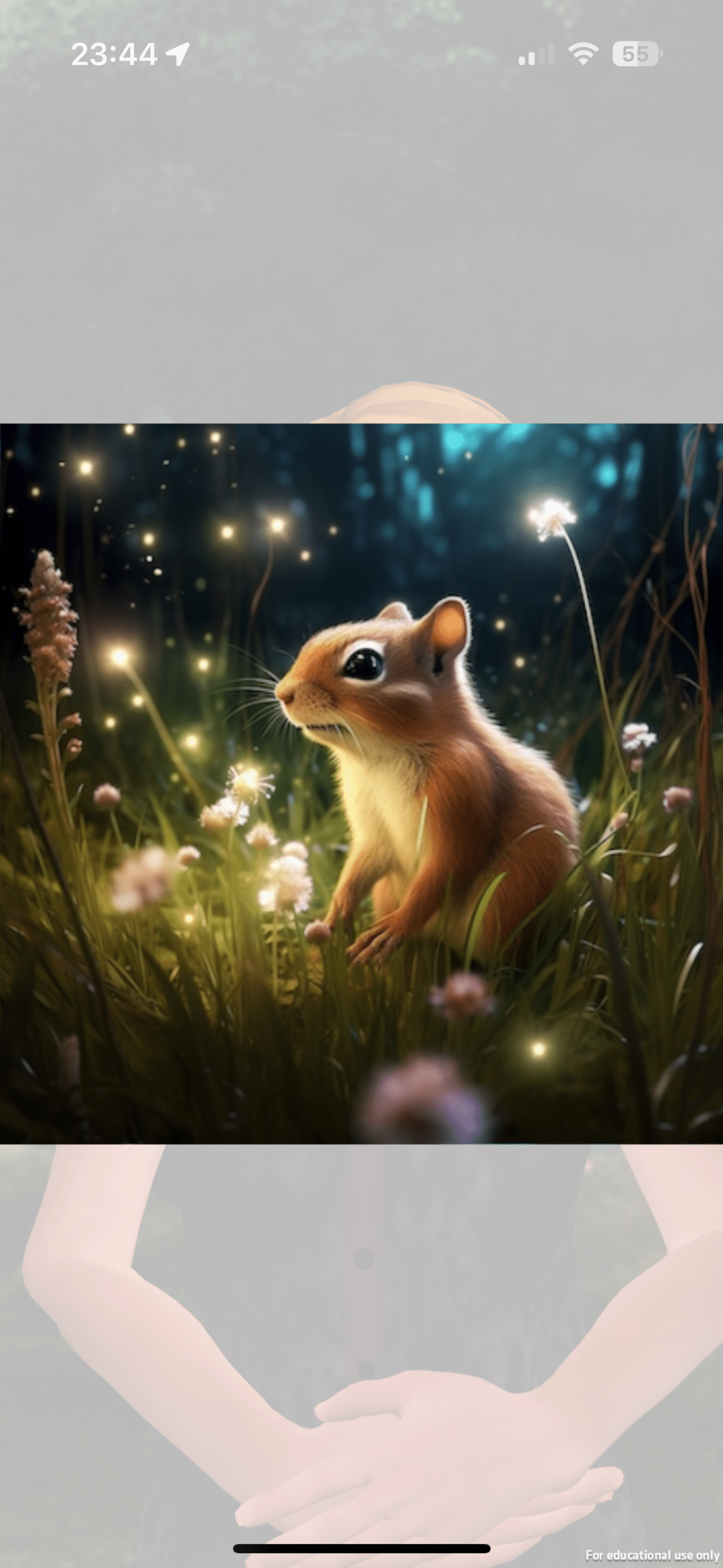} 
    \caption{Forest Exploration}
    \label{fig:squirrel}
  \end{subfigure}
  \caption{The ClaraEdu Interface. (a) The welcome screen frames the content source and offers users a choice to engage or verify information directly. (b) Dr. Clara delivers the non-gamified pedagogical version for parents and adolescents. (c-e) The gamified narrative path features a narrator introducing the game, character customization options, and forest creatures posing HPV-related riddles.}
  \label{fig:teaser}
  \Description{Five mobile phone screenshots showing the ClaraEdu app. First, a welcome screen with CDC attribution and options to talk to Clara or view the CDC page. Second, Dr. Clara, a Black female doctor in a white coat in a medical office. Third, a narrator character with brown hair against a pink sunset, with dialogue options about saving a village. Fourth, a role selection screen with options like Adventurer, Scientist, Mage, Warrior, and Healer. Fifth, an illustrated squirrel in a magical forest with glowing lights.}
\end{teaserfigure}

\maketitle

\section{Introduction}
Vaccine hesitancy, accelerated by online misinformation, critically impacts uptake of the Human Papillomavirus (HPV) vaccine, which, despite proven safety and ability to prevent over 90\% of HPV-related cancers, reaches only half of eligible adolescents in the United States \cite{dube2013vaccine, sonawane2020parental, kamolratanakul2021human, cdc_hpv_vaccination_2024}. HPV vaccination typically requires assent from children during pre-adolescence—yet adolescents, who could participate in clinical decision-making if appropriately empowered, often lack the health literacy to understand medical information \cite{holman2014barriers} and are treated as passive recipients rather than patients with developing levels of agency \cite{dedding2015revealing, gowda2012understanding}. Additionally, giving adolescents a voice requires mechanisms for their perspectives to reach decision-makers, even when they hesitate to speak up.

Embodied Conversational Agents (ECAs)—animated virtual characters—have demonstrated success in health education by providing judgment-free environments for information exchange \cite{bickmore2009using, gogoi2022computer}. However, few systems have attempted to mediate the parent–child dyad simultaneously. Designing for this dual audience requires balancing parental authority with the adolescent’s evolving need and capacity for agency and age-appropriate information delivery.

This paper presents the design and pilot evaluation of \textit{ClaraEdu}, a mobile intervention that gives adolescents a voice in HPV vaccination decisions alongside their parents. The system offers a "dual-path" design: parents interact with a simulated physician, while adolescents choose between a standard physician or a gamified narrative experience. We report findings from a clinic-based pilot study with 21 parent–adolescent dyads. 
We hypothesize that: \textbf{H1} (Satisfaction \& Attitude): Both parents and adolescents will report satisfaction and attitude scores, regarding the app and conversational agent, significantly above neutral; and \textbf{H2} (Exploratory Efficacy): Both parents and adolescents will demonstrate positive trends in HPV knowledge, attitude, and vaccination intention from baseline to post-intervention.

\section{Related Work}
\subsection{ECAs in Health Education}
ECAs have been increasingly utilized in health interventions, combining empathetic nonverbal cues like hand gestures and facial expressions with spoken dialog to facilitate health behavior change, particularly for individuals with limited health literacy \cite{kocaballi2022design, wang2015acceptability, tudor2020conversational, bickmore2009using, murali2019health, bickmore2013randomized, bickmore2015automated, steenstra2023changing}. These agents have been successfully deployed across diverse health domains, such as psychotherapy \cite{10.1145/3706599.3720086, provoost2017embodied} and substance use disorder treatment \cite{olafsson2020towards, zhou2017relational, bickmore2020substance, steenstra2024virtual}. More recently, ECAs have expanded into vaccination education, with web-based tailored interventions delivered by ECAs positively impacting informed decision-making around HPV vaccination by decreasing decisional conflict and increasing uptake \cite{pot2017effectiveness}, and pilot studies finding that computer-animated relational agents were more effective at motivating discussions about HPV vaccination than traditional brochures \cite{gogoi2022computer}. These digital interventions have shown promise in outperforming traditional methods like pamphlets, which have had limited effectiveness in promoting vaccination and enhancing knowledge \cite{merchant2020comparison, thomas2018interventions, tan2020motivational, murali2022training}.

\subsection{Digital Health for Family Vaccination Decision-Making}
The landscape of digital tools for family vaccination spans from simple reminders to comprehensive applications, with a central paradox emerging: while low-tech interventions like SMS reminders consistently demonstrate a median 6-12 percentage point increase in vaccination uptake by solving logistical barriers \cite{cpstf2015increasing}, a systematic review of 25 childhood vaccination apps found "little evidence" that they improve uptake, even when well-received by users \cite{de2020use}. For example, applications like \textit{Vaccipack}—co-designed with adolescents and parents to promote HPV vaccination—report high acceptability, but the direct link to improved vaccination rates remains unclear \cite{teitelman2020vaccipack}. This gap reflects the difficulty of influencing complex drivers of parental vaccine hesitancy, particularly misinformation encountered online \cite{kolff2018use}. Research has consequently shifted toward more targeted, relational, and persuasive approaches \cite{olson2020addressing, feldman2022smartphone, koskan2024protocol}. Our work builds on this trajectory by using an ECA to deliver a persuasive, empathetic, and tailored conversational experience supporting shared decision-making for both parents and adolescents.

\subsection{Health Behavior Change in HCI}
Designing technologies to support health behavior change is a core HCI focus, with a central challenge of understanding not just if an intervention works, but how, for whom, and in what context \cite{klasnja2011evaluate, klasnja2017toward}. HCI research confirms that one-size-fits-all approaches are ineffective, as individuals have highly personal needs and motivations when changing health behaviors \cite{paay2015understanding, oyebode2021tailoring}, leading to systems supporting personalized goal setting and contextual notifications \cite{zhu2025systematic, chen2024investigating}. These challenges are amplified for sensitive health topics, where family communication is hindered by online misinformation and privacy concerns, particularly for adolescents \cite{konstantinou2025behavior, dewan2024teen}. HCI has explored technologies facilitating Shared-Decision Making, with conversational agents showing promise in making complex information accessible \cite{hao2024advancing, samiee2025general}, while game-based learning improves knowledge and motivation on sensitive health topics for younger audiences \cite{haruna2018improving, schoech2013gamification}.

\section{System Design: The ClaraEdu Application}
Building directly upon the outcomes of our prior dyadic and gamified interventions, we developed \textit{ClaraEdu}, a mobile application for iOS and Android designed to facilitate HPV vaccination decision-making. The system uses ECAs tailored for both parent and adolescent users (Figure~\ref{fig:dr-clara}). It uses a 3D rendering engine to create animated characters with synchronized speech and linguistically-appropriate nonverbal behaviors generated via the BEAT engine \cite{cassell2001beat}. Dialogue is managed through a hierarchical transition network, with users interacting via multiple-choice options. All content was developed in collaboration with and verified by clinical collaborators specializing in HPV vaccination and health communication.  

\textbf{Phase 1: Rapport, Permission, and Framing.}
The initial moments of a health interaction are critical for establishing trust and setting a collaborative tone \cite{dang2017building}. To address this, the application opens with a clear framing of the content source (CDC recommendations) (Figure~\ref{fig:welcome}), allowing users to verify the material directly. Furthermore, to empower users and reinforce their autonomy over health decisions \cite{hutton2023digital}, the agent explicitly 
requests permission from the user to discuss the topic. 

\textbf{Phase 2: Staging and Path Tailoring.}
The system uses the Transtheoretical Model (TTM) to match intervention intensity with a user's readiness to change \cite{prochaska1997transtheoretical}, preventing disengagement by avoiding unsolicited advice for those not yet ready to hear it. To operationalize this, the agent assesses the user's 'Stage of Change' via a direct inquiry regarding vaccination intent and tailors their educational path accordingly. For example, adolescents identified in the \textit{Contemplation} stage—those considering vaccination but hesitant—the system integrates our prior work on a narrative game as an optional path (Figure~\ref{fig:narrator}) \cite{steenstra2024_game}. Developed using an expert-in-the-loop methodology where clinicians vetted LLM-generated assets for medical accuracy, this path invites adolescents to customize a hero avatar (Figure~\ref{fig:role-select}) and navigate a fantasy world. HPV and vaccination educational content is embedded directly into gameplay mechanics; for example, to unlock new areas, adolescents must correctly solve riddles regarding vaccine efficacy posed by animated forest creatures (Figure~\ref{fig:squirrel}). This design choice leverages our finding that narrative-based gamification significantly increases adolescent engagement and entertainment compared to pedagogical approaches \cite{steenstra2024_game}, making it an ideal medium for capturing the attention of undecided adolescents and motivating consideration of behavior \cite{hamari2014does, haruna2018improving}.

\textbf{Phase 3: Tailored Education and Engagement.}
To ensure health education is accessible to individuals with varying health literacy levels \cite{bickmore2009using}, the agent uses plain language, avoids jargon, and structures information clearly. While both the parent and child versions cover identical substantive content—transmission, cancer risks, vaccine efficacy, dosing, and safety—the presentation is adapted for each audience. The parent version focuses on detailed efficacy data, while the adolescent version utilizes age-appropriate analogies. For the younger audience, the optional gamified path leverages narrative learning principles to improve knowledge retention and motivation \cite{hamari2014does, haruna2018improving}. Both versions utilize comprehension checks and Q\&A menus to maintain active engagement.

\textbf{Phase 4: Empathetic Exploration of Hesitancy.}
For users who express disinterest or resistance (Precontemplation Stage), continuing with unsolicited information risks alienating them entirely. Drawing on Motivational Interviewing principles \cite{miller2012motivational}, the system adopts a collaborative, non-judgmental stance. The agent explores hesitancy by asking users to identify specific concerns, validates their perspective (e.g., acknowledging protective instincts), and attempts to elicit "change talk" by asking them to articulate their own reasons favoring vaccination. The module culminates with a readiness ruler (1-10 scale) as a way to reinforce intrinsic motivation \cite{miller2012motivational}.

\textbf{Phase 5: Facilitating Dyadic Communication.}
Effective family health decisions require productive parent–adolescent communication, yet topics such as vaccination can be difficult to discuss \cite{anderson2025evolving}. Prior work shows that parents and adolescents often differ in their vaccine readiness and that adolescents’ intentions can influence parental decisions \cite{oudin2024shared}, and that adolescents frequently feel sidelined in vaccine decisions yet desire more information and a greater role \cite{mitchell2022adolescents}. Building on this evidence and our prior dyadic agent work \cite{steenstra2023changing}, \textit{ClaraEdu} uses Shared Decision Making principles to honor both parental guidance and the adolescent's developing autonomy. The agent coaches both parties separately on how to engage constructively. In line with TTM and MI–based recommendations to match guidance to readiness \cite{prochaska1997transtheoretical,miller2012motivational}, parents who already intend to vaccinate receive brief guidance on initiating clear, informational conversations, whereas more hesitant parents are encouraged toward a more open, exploratory dialogue that elicits their child's questions and concerns. Similarly, adolescents who are classified as ready to vaccinate (via the Stage of Change assessment in Phase~2) are coached to express decisions clearly, while those who are unsure are guided on how to voice their concerns \cite{mitchell2022adolescents}. All users receive specific coaching on how to use the upcoming clinical appointment for further discussion \cite{blanchard2019pre}.

\textbf{Phase 6: Resolve Barriers to Care.}
Translating intention into action requires addressing practical obstacles that can prevent follow-through. The system proactively identifies logistical barriers—such as transportation, cost, and scheduling concerns—and provides targeted solutions to resolve them. To address informational gaps and residual uncertainties, the system allows both parents and adolescents to flag specific questions about safety, efficacy, and recommendations. These questions are compiled and transmitted to the clinic staff prior to the visit. For adolescents, this provides a concrete voice mechanism—questions they hesitate to ask aloud still reach the provider.

\begin{table*}[t]
\centering
\small
\begin{tabular}{l c c c c c c}
\toprule
& \multicolumn{3}{c}{Parent} & \multicolumn{3}{c}{Child} \\
\cmidrule(lr){2-4} \cmidrule(lr){5-7}
Item (Likert Scale – Higher is Better) & Mean & SD & $p$ & Mean & SD & $p$ \\
\midrule
How easy was it to use the app? (1-7) & 6.82 & 0.39 & $<.05$ & 6.50 & 0.53 & $<.05$ \\
How satisfied are you with the animated character? (1-7) & 5.59 & 1.77 & $<.05$ & 5.13 & 2.23 & $0.20$ \\
How natural was your conversation with the animated character? (1-7) & 4.88 & 2.20 & $0.12$ & 4.63 & 2.33 & $0.47$ \\
How much do you feel the animated character cares about you? (1-7) & 4.41 & 2.55 & $0.52$ & 5.13 & 1.46 & $0.07$ \\
How would you characterize your relationship with the animated character? (1-7) & 3.35 & 2.42 & $0.29$ & 3.75 & 2.55 & $0.79$ \\
How much do you trust the animated character? (1-7) & 3.94 & 2.44 & $0.92$ & 4.00 & 2.39 & $1.00$ \\
How much would you like to continue working with the animated character? (1-7) & 5.24 & 1.89 & $<.05$ & 4.75 & 2.31 & $0.39$ \\
How much did you like the animated character? (1-7) & 5.35 & 2.00 & $<.05$ & 4.50 & 2.67 & $0.61$ \\
The animated character was a good way for my child to learn about HPV. (1-7) & 6.25 & 1.04 & $<.05$ & - & - & - \\
The animated character changed my attitude about HPV vaccination. (1-5) & 3.71 & 1.05 & $<.05$ & 2.75 & 1.39 & $0.63$ \\
I learned a lot about HPV from the animated character. (1-5) & 4.71 & 0.47 & $<.05$ & 3.25 & 1.58 & 0.69 \\
My child enjoyed interacting with the animated character. (1-5) & 3.71 & 1.70 & $0.31$ & - & - & - \\
I felt actively involved in the decision-making process. (1-5) $^{\dagger}$ & - & - & - & 3.60 & 0.52 & $<.05$ \\
Learning about HPV from the animated character(s) is fun. (0-3) $^{\dagger}$ & - & - & - & 2.09 & 0.92 & $0.11$ \\
\bottomrule
\multicolumn{7}{l}{\footnotesize $^{\dagger}$ Composite measure derived from multiple survey items.}
\end{tabular}
\caption{Combined satisfaction and attitude ratings for parents and children. One-sample t-tests demonstrate scores significantly different from neutral.}
\label{tab:combined_satisfaction}
\end{table*}

\begin{table*}[t]
  \centering
  \label{tab:{descriptives}}
  \setlength{\tabcolsep}{4pt} 
  \begin{tabular}{lcccccccccccc}
    \toprule
    & \multicolumn{4}{c}{\textbf{CONTROL}}
    & \multicolumn{4}{c}{\textbf{PARENT}}
    & \multicolumn{4}{c}{\textbf{CHILD}} \\
    \cmidrule(lr){2-5} \cmidrule(lr){6-9} \cmidrule(lr){10-13}
    \textbf{Measure}
    & \multicolumn{2}{c}{Parent $\Delta$} & \multicolumn{2}{c}{Child $\Delta$}
    & \multicolumn{2}{c}{Parent $\Delta$} & \multicolumn{2}{c}{Child $\Delta$}
    & \multicolumn{2}{c}{Parent $\Delta$} & \multicolumn{2}{c}{Child $\Delta$} \\
    \cmidrule(lr){2-3} \cmidrule(lr){4-5}
    \cmidrule(lr){6-7} \cmidrule(lr){8-9}
    \cmidrule(lr){10-11} \cmidrule(lr){12-13}
    & M & SD & M & SD
    & M & SD & M & SD
    & M & SD & M & SD \\
    \midrule
    HPV Knowledge &  0.75 &  1.89 & – & – &  3.00 &  3.20 & – & – &  2.25 &  1.83 & – & – \\
    HPV Attitude & -0.29 & 1.09 & 0.40 & 0.43 & 0.10 & 1.00 & -0.16 & 0.68 & -0.43 & 1.50 & 0.30 & 0.71 \\
    HPV Vaccine Intent  & 0.00 & 1.41 & 0.25 & 1.26 & 0.44 & 0.53 & 0.33 & 1.12 & 1.00 & 1.07 & 1.25 & 1.28 \\
    \bottomrule
  \end{tabular}
  \caption{HPV Knowledge, Attitude, and Intent for parents and children. Means (M) and standard deviations (SD) of pre-post changes ($\Delta$) by condition and respondent.}
  \label{tab:outcomes}
\end{table*}

\section{Evaluation Study}
To evaluate the efficacy of the HPV vaccination promotion intervention we planned and began recruiting for a randomized controlled trial, with a target recruitment of 875 parent–adolescent dyads randomized to a non-intervention control group (CONTROL), a group in which only the parent had access to the intervention (PARENT), or a group in which both the parent and adolescent had access (DYAD). Primary outcomes included actual HPV vaccination uptake within 16 months, along with a range of secondary measures collected just before and after a pediatric clinic visit during which time the intervention was made available. Secondary measures included HPV knowledge, HPV attitudes, and satisfaction with and attitudes towards the intervention app and conversational agent. However, five months into the trial, the NIH informed us that the grant funding the project was terminated.\footnote{The termination letter stated "It is the policy of NIH not to prioritize research activities that focuses gaining scientific knowledge on why individuals are hesitant to be vaccinated and/or explore ways to improve vaccine interest and commitment."} 
At this point, we had recruited 50 parent–adolescent dyads, but only 21 had completed pre- and post-visit measures. Here, we report these partial results as pilot data for a clinical trial we hope to complete in the future. Our IRB approved the trial, and participants were compensated for their time. 

\textbf{Measures.}
HPV Knowledge (10 true/false/don't know items, scored as proportion correct) and HPV Attitudes (7 items on 5-point Likert scales) were assessed using items adapted from the Knowledge and Attitude toward HPV Questionnaire \cite{lopez2022hpv}, with additional knowledge items from Sallam et al. \cite{sallam2019dental} and attitude items from the Knowledge and Acceptability of Papillomavirus Vaccines in Parents of Adolescents in Spain study \cite{lopez2022hpv}. Vaccination Intention used single-item measures (5-point scale) adapted from the National Immunization Survey-Teen\footnote{https://www.cdc.gov/nis/php/datasets-teen/index.html}. Satisfaction and attitude, regarding the app and conversational agent, were assessed using items from the General Agent Rating scale \cite{olafsson2021more} and Engagement versus Disaffection with Learning scale \cite{skinner2009engagement}. Lastly, for adolescent decision-making on HPV vaccination, items were taken from the HPV Adolescent Vaccine Intervention Questionnaire \cite{forster2017development}.

\textbf{Participants.}
Participants were recruited from a pediatric clinic at a safety-net hospital prior to a well-child visit. Eligibility criteria included: parent/guardian $\geq$18 years old, adolescent aged 8-12 years, English fluency, adolescent had received zero or one (of two) HPV vaccine doses, smartphone access, and an upcoming well-child visit. 

\textbf{Procedure.}
Participants completed baseline surveys prior to app access and post-visit surveys within two weeks following their clinic appointment. PARENT and CHILD participants received download links for the app and were instructed to interact with it prior to their well-child clinic visit.

\textbf{Results.}
Of the 21 dyads who completed all measures, parents were aged 29-54 ($M=39, SD=7.69$), all female, and 66.7\% Black, and adolescents were aged 9-12 ($M=9.25, SD=1.25$), 61.9\% female, and 66.7\% Black. Four (4) dyads were randomized to CONTROL, 9 to PARENT only, and 8 to CHILD. 

\textbf{Satisfaction and Attitude (H1).} Results from the post-intervention survey support H1, indicating that the intervention was well-received by both audiences. As detailed in Table~\ref{tab:combined_satisfaction}, parents rated the ease of use ($M=6.82$) and the agent's suitability as a learning tool for their child ($M=6.25$) significantly above the neutral midpoint. While parents reported moderate levels of trust ($M=3.94$) and relationship formation ($M=3.35$) with the agent, they expressed a strong desire to continue working with the character ($M=5.24$). Adolescents similarly rated ease of use highly ($M=6.50$) and reported feeling that the agent ``cared'' about them ($M=5.13$). Notably, adolescents reported a high sense of agency, rating their involvement in the HPV vaccine decision-making process significantly above neutral ($M=3.60$). Regarding the content format, three (3, 37.5\%) of the adolescents in the CHILD condition chose to use the gamified version over the standard doctor interaction.

\textbf{Exploratory Efficacy: HPV Knowledge, Attitude, and Intent (H2).} No statistically significant between-group differences were found due to the small number of participants resulting from the trial's early termination. However, distinct trends emerged supporting H2 (Table~\ref{tab:outcomes}). In terms of HPV knowledge, parents in both intervention arms demonstrated notable improvements (PARENT $\Delta=3.00$; CHILD $\Delta=2.25$) compared to the control group ($\Delta=0.75$). However, HPV attitude changes were mixed across all groups. The most distinct pattern emerged regarding HPV vaccination intention. While the PARENT-only condition showed modest gains in parental intent ($\Delta=0.44$), the CHILD condition—where the adolescent participated—showed the largest increases in intent for both parents ($\Delta=1.00$) and adolescents ($\Delta=1.25$).

\section{Discussion}
Our findings offer preliminary support for the efficacy of giving adolescents a voice in health prevention technologies. Supporting H1, the high satisfaction and attitude scores demonstrate that a dual-audience ECA framework is an acceptable and engaging modality for delivering sensitive health information to families. Supporting H2, the exploratory data suggest that this engagement translates into cognitive and motivational gains. Most critically, the data reveal a ``dyadic advantage'' regarding vaccination intention: intent scores increased most sharply in the condition where both the parent and child interacted with the system (Table~\ref{tab:outcomes}). This aligns with prior research on Shared Decision-Making, suggesting that when adolescents are treated as stakeholders with agency rather than passive subjects, they become active partners in the health decision process, potentially reducing parental hesitation and logistical friction \cite{miller2019decision, miller2023adolescent, wyatt2015shared}.

Several design insights emerged. Designing for dyads rather than individuals required the agent to prepare both parties for subsequent conversation with each other, not just educate them separately, addressing a gap in individual-focused interventions \cite{anderson2025evolving}. The question transmission feature exemplifies this: adolescents can flag concerns that are delivered directly to providers, ensuring their voice reaches the clinical encounter even if they remain silent during the visit. Lastly, offering the gamified path as an option respected user preferences while maintaining content fidelity, supporting personalization approaches for younger audiences \cite{haruna2018improving}.

Several limitations constrain interpretation. Most critically, small and unbalanced samples resulting from study termination severely limited statistical power; trending patterns across multiple outcomes suggest effects that adequately powered studies might detect. The study also lacked long-term follow-up and behavioral verification through medical records—future work must assess vaccination uptake, not just intentions.

\section{Conclusion}
We presented \textit{ClaraEdu}, a dual-audience ECA designed to support shared decision-making for HPV vaccination. Pilot results indicate high satisfaction and suggest that providing adolescents with direct, age-appropriate education tools may enhance vaccination intention within the family unit. Future work will validate these dyadic mechanics in a fully powered clinical trial. 

\begin{acks}
This work was supported by the National Institutes of Health, National Cancer Institute (until it was terminated), under Grant R01CA273208.
\end{acks}

\balance 

\bibliographystyle{ACM-Reference-Format}
\bibliography{bibliography}

\end{document}